\documentclass[envcountsect,runningheads]{llncs}
\usepackage[utf8]{inputenc}
\usepackage{mathtools}
\usepackage{amsmath,amssymb,cancel,xcolor,tikz,bussproofs,multicol,hyperref,multirow,paralist}
\usepackage{todonotes}

\usepackage{subcaption}

\usepackage{mathpartir}

\usepackage{algorithm}
\usepackage[noend]{algpseudocode}
\usepackage[misc,geometry]{ifsym}

\usepackage{tikz}

\def\orcidID#1{\href{http://orcid.org/#1}{\raisebox{-1.25pt}{\includegraphics{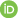}}}}

\newcommand{\vampire}{\textsc{Vampire}}

\newcommand{\CommentedOut}[1]{}
\newcommand{\SR}{\texttt{SR}}

\renewcommand{\paragraph}[1]{\par\smallskip\noindent\textbf{#1}}

\newcommand{\solver}{\mathit{solver}}
\newcommand{\ms}{\mathit{ms}}  
\newcommand{\NoSubsumptionResolution}{\texttt{NoSubsumptionResolution}}
\newcommand{\NoSubsumption}{\texttt{NoSubsumption}}
\newcommand{\Subsumption}{\texttt{Subsumed}}
\newcommand{\SAT}{\texttt{SAT}}

\title{SAT-Based Subsumption Resolution}
\titlerunning{SAT-Based Subsumption Resolution}

\author{%
    Robin Coutelier\inst{1}\textsuperscript{(\Letter)}
    \and
    Laura Kovács\inst{2}\orcidID{0000-0002-8299-2714}
    \and
    Michael Rawson\inst{2}\orcidID{0000-0001-7834-1567}
    \and
    Jakob Rath\inst{2}\orcidID{0000-0003-0346-6749}
}
\institute{%
    U. Li\`ege, Li\`ege, Belgium\\
    \texttt{robin.coutelier@student.uliege.be}
    \and
    TU Wien, Vienna, Austria
}
\authorrunning{Robin Coutelier, Laura Kovács, Michael Rawson, and Jakob Rath}
\date{}

\begin{document}

\maketitle

\begin{abstract}
Subsumption resolution is an expensive but highly effective simplifying
inference for first-order saturation theorem provers.
We present a new SAT-based reasoning technique for subsumption resolution,
without requiring radical changes to the underlying saturation algorithm.
We implemented our work in the theorem prover \vampire,
and show that it is noticeably faster than the state of the art.
\begin{tikzpicture}[remember picture, overlay]%
    \node[anchor=south,align=left,text width=11cm,minimum height=3cm] at (current page.south) {
        This version of the contribution has been accepted for publication,
        after peer review but is not the Version of
        Record and does not reflect post-acceptance improvements, or any
        corrections. The Version of Record is available online at:
        \url{https://dx.doi.org/10.1007/978-3-031-38499-8_11}
    };%
\end{tikzpicture}%
\end{abstract}

\section{Introduction}%
\label{sec:introduction}

Saturation-based proof search is a popular approach to first-order theorem proving~\cite{DBLP:conf/cav/KovacsV13,E19,Spass09}.
In addition to efficient inference systems~\cite{Rubio01,BG94}, saturation
provers also implement \emph{redundancy elimination} to reduce the size of the search space.
Redundancy elimination deletes clauses from the search space by showing them to be logical consequences of other (smaller) clauses, and therefore redundant.
However, checking whether a first-order formula
is implied by another first-order formula is undecidable, and so eliminating redundant clauses
is in general undecidable too. In practice, saturation systems apply cheaper conditions for redundancy elimination, such as removing equational tautologies by congruence closure or deleting subsumed clauses by establishing multiset inclusion.
Recently, SAT solving has been applied to efficiently detect and remove subsumed clauses~\cite{rath2022first}.
\emph{We extend SAT-based reasoning in first-order theorem proving to a combination of subsumption and resolution, \textbf{subsumption resolution}~\cite{DBLP:books/el/RV01/BachmairG01} (Section~\ref{sec:sat-based-approach}).}

Both subsumption and subsumption resolution are NP-complete~\cite{matching-np-complete}.
To improve efficiency in practice, we  (i) encode  subsumption resolution as  SAT formulas over (match) set constraints (Section~\ref{sec:subsumption-resolution}) and (ii) directly integrate CDCL SAT solving for checking subsumption resolution in  first-order theorem proving (Section~\ref{sec:subsumption-resolution-in-vampire}). 
We implement our approach in the theorem prover \vampire{}~\cite{DBLP:conf/cav/KovacsV13}, improving the state-of-the-art in first-order reasoning (Section~\ref{sec:experiments}).

\paragraph{Related Work.}
Subsumption and subsumption resolution are some of the most powerful and frequently used redundancy criteria in saturation-based provers. Subsumption resolution is supported as \emph{contextual literal cutting} in~\cite{E19}, along with efficient approaches for detecting multiset inclusions among clauses~\cite{DBLP:conf/cav/KovacsV13,Spass09,schulz2013simple}. 
Special cases of {unit deletion} as a by-product of subsumption tests are also proposed in~\cite{Tammet:1998}.
Much attention has been given to refinements of \emph{term indexing}~\cite{Tammet:1998,schulz2013simple}
to drastically reduce the set of candidate clauses checked for subsumption. Recently, these approaches have been complemented by SAT solving~\cite{rath2022first}, reducing subsumption checking to SAT. Our work generalises this approach by solving for both subsumption and subsumption resolution via SAT. 

SAT solvers have been applied widely to first-order theorem proving, including but not limited to AVATAR~\cite{DBLP:conf/cav/Voronkov14}, instance-based methods~\cite{inst-gen}, heuristic grounding~\cite{E19}, global subsumption~\cite{global-subsumption} and combinations thereof~\cite{uses-of-sat}, but using SAT solvers for classical subsumption methods is under-explored. To the best of our knowledge, SAT solving for subsumption resolution has so far not been addressed in the landscape of automated reasoning.

\section{Illustrative Examples and Main Contributions}%
\label{sec:motivating-example}
Let us illustrate a few challenges of subsumption resolution, which motivate our approach to solving it (Section~\ref{sec:sat-based-approach}). 
Given a pair of clauses~$L$ and~$M$, denoted as $(L,M)$, 
the problem is to decide whether~$M$ can be simplified by~$L$ via a special case of logical consequence.
In Figure~\ref{fig:Examples} we show examples  where it is not obvious for which pairs $(L_i,M_i)$ subsumption resolution can be applied.
\begin{figure}[!ht]
\begin{center}
    \begin{tabular}{|c|c|}
        \hline
        \parbox{6cm}{
            \begin{align*}
                L_1 & := p(x_1, x_2) \lor p(f(x_2), x_3) \\
                M_1 & := p(g(y_1), c) \lor \neg p(f(c), e)
            \end{align*} 
        } &
        \parbox{6cm}{
            \begin{align*}
                 L_2 & := p(x_1) \lor q(x_2) \\
                 M_2 & := \neg p(y) \lor \neg q(c)
             \end{align*}
         }\\
         \hline
         \parbox{6cm}{
             \begin{align*}
                 L_3 & := p(x_1) \lor q(x_1, x_2) \lor \neg p(x_2) \\
                 M_3 & := \neg p(y) \lor q(y,y)
             \end{align*} 
         } &
         \parbox{6cm}{
             \begin{align*}
                 L_4 & := p(x_1) \lor q(x_2) \lor r(x_3) \\
                 M_4 & := \neg p(y_1) \lor q(c)
             \end{align*}
         }\\
         \hline
    \end{tabular}
\end{center}
\caption{Illustrative examples.\label{fig:Examples}}
\end{figure}

In fact, subsumption resolution can only be applied to $(L_1, M_1)$.
Later, we show  how our approach determines that $M_1$ can be
shortened in the presence of $L_1$ (Example~\ref{ex:SubsRes}),
but also how the remaining pairs
cannot apply subsumption resolution (Examples~\ref{ex:uniqueness}, \ref{ex:coherence}, and \ref{ex:SAT:pruning}).
For example, $(L_4,M_4)$ is
filtered by \emph{pruning} to bypass the SAT routine altogether.

\paragraph{Our Contributions.}
\begin{enumerate}
    \item We cast the problem of subsumption resolution over pairs of first-order formulas $(L,M)$ as a SAT problem
    (Theorem~\ref{thm:subsumption-resolution-constraints}), ensuring any instance of subsumption resolution is a model of this SAT problem. 
    \item We tailor encodings of subsumption resolution (Sections~\ref{sec:direct:SR}--\ref{sec:indirect:SR}) for effective SAT-based subsumption resolution (Algorithm~\ref{alg:sat-subsumption-resolution}). 
    \item  We integrate our approach into the saturation loop, solving for subsumption and subsumption resolution simultaneously (Section~\ref{sec:subsumption-resolution-in-vampire}).
    \item We implement our work in the theorem prover \textsc{Vampire} and showcase our practical gains in first-order proving (Section~\ref{sec:experiments}). 
\end{enumerate}

\section{Preliminaries}%
\label{sec:preliminaries}

We assume familiarity with first-order logic with equality. We include standard Boolean connectives and quantifiers in the language, and the constants $\top, \bot$ for truth and falsehood.
We use $x,y,z$ for first-order variables, $c,d,e$ for constants, $f, g$ for functions, $p, q, r$ for atoms, $l,m$ for literals, and $L, M$ for clauses, all potentially with indices.
If $L$ is a clause $l_1\vee \ldots \vee l_n$, we sometimes consider it as a multiset of its literals $l_i$, and write $|L|$ for its cardinality (i.e. the number $n$ of literals in $L$). The empty clause is denoted $\square$.
Free variables are universally quantified.
An expression $E$ is a term, atom, literal, clause, or formula. 

\paragraph{Substitutions and matches.}
A substitution~$\sigma$ is a (partial) mapping from variables to terms.
The result of applying a substitution~$\sigma$ to an expression~$E$ is denoted $\sigma(E)$
and is the expression obtained by simultaneously replacing each variable~$x$ in~$E$ by~$\sigma(x)$.
For example, the application of $\sigma \coloneqq \{ x \mapsto f(c) \}$
to the clause $L \coloneqq \{ p(x), q(x, y) \}$
yields $\sigma(L) = \{ p(f(c)), q(f(c), y) \}$. 
Note that $\sigma(L)$ is a logical consequence of $L$.

A  \emph{matching substitution}, in short a \emph{match}, between literals $l$ and $m$
is a substitution~$\sigma$ such that $\sigma(l) = m$.
For example, the match of $p(x)$ onto $p(f(c))$ is $\{x\mapsto f(c)\}$.
Two matches are \emph{compatible} and can be combined in the same substitution
iff they do not assign different terms to the same variable. 
For example, the substitutions $\{x\mapsto f(c), y\mapsto g(d)\}$ and
$\{x\mapsto f(c), z\mapsto h(e)\}$ are compatible,
but $\{x\mapsto f(c)\}$ and $\{x\mapsto g(c)\}$ are not.

\paragraph{Saturation and redundancy.}
Many first-order systems apply the superposition calculus~\cite{BG94} in a saturation loop~\cite{Rubio01}.
Given an input set $F$ of clauses,
saturation iteratively derives logical consequences and adds them to $F$.
By soundness and completeness of superposition, 
if $\square$ is derived the system can report unsatisfiability of $F$; if $\square$ is not encountered and no further clauses can be derived, the system reports satisfiability of $F$.

Saturation is more efficient when $F$ is as small as possible. For this reason, saturation-based provers also employ \emph{simplifying} inferences. Simplifying inferences
reduce the number or size of clauses in $F$.
This is formalised using the following notion of \emph{redundancy}:
a ground clause~$M$ is redundant in a set of ground clauses~$F$
if~$M$ is a logical consequence of clauses in~$F$ that are strictly smaller than~$M$
w.r.t.\ a fixed simplification ordering $\succ$.
A non-ground clause $M$ is redundant in a set of clauses~$F$
if each ground instance of~$M$ is redundant in the set of ground instances of~$F$.
If $M$ is redundant in $F$, then $M$ can be removed from $F$ while retaining completeness.

\paragraph{Subsumption.} A clause $L$ \emph{subsumes} a distinct clause $M$ iff there is a substitution $\sigma$ such that
\begin{equation}\label{eq:subs}
  \sigma(L)\subseteq_M M
\end{equation}
where $\subseteq_M$ denotes multiset inclusion. We also say that $M$ is \emph{subsumed} by $L$. Note that subsumed clauses are redundant.

Removing subsumed clauses $M$ from the search space $F$ is implemented through a simplifying rule,  checking condition~\eqref{eq:subs} over pairs of clauses $(L,M)$ from $F$. Matches between every literal in $L$ to some literal in $M$ are checked; if a compatible set of matches is found, then $M$ can be removed from $F$.

\paragraph{Subsumption resolution.}
Subsumption resolution aims to remove one redundant literal
from a clause. Clauses $M$ and $L$ are said to be the main and side premise of subsumption resolution, respectively, iff there is a substitution $\sigma$, a set of literals $L'\subseteq L$ and a literal $m'\in M$ such that
\begin{equation}
  \label{eq:subsumption-resolution-definition}
  \sigma(L') = \{\neg m'\}
  \quad\text{and}\quad
  \sigma(L\setminus L') \subseteq M\setminus \{m'\}
  .
\end{equation}
If so, $M$ can be replaced by $M\setminus \{m'\}$.
Subsumption resolution is hence the rule
\begin{prooftree}
  \AxiomC{$L$}
  \AxiomC{$\cancel{M}$}
  \LeftLabel{(\SR)~~~}
  \BinaryInfC{$M \setminus \{m'\}$}
\end{prooftree}
 
We indicate the deletion of a clause $M$ by drawing a line through it ($\cancel{M}$),
and we refer to the literal $m'$ of $M$ as the \emph{resolution literal} of \SR.
Intuitively, subsumption resolution is binary resolution followed by subsumption of one of its premises by the conclusion.
However, by combining two inferences into one it can be treated as a simplifying inference,
which is advantageous from the perspective of proof search dynamics.

\begin{example}%
  \label{ex:SubsRes}
  Consider $L_1,M_1$ of Figure~\ref{fig:Examples}. 
Subsumption resolution is applied
by using the substitution
$\sigma \coloneqq \{x_1 \mapsto g(y_1), x_2 \mapsto c, x_3 \mapsto e\}$.
Note that  $\sigma(L_1) = p(g(y_1), c) \lor p(f(c), e)$. 
$\sigma(L_1)$ and $M_1$ can be resolved to obtain $p(g(y_1), c)$.
The clause $p(g(y_1), c)$  subsumes $M_1$, since it is a sub-multiset of $M_1$. We have 
\[
    \inferrule
    {p(x_1, x_2) \lor p(f(x_2), x_3) \\ \cancel{p(g(y_1), c) \lor \neg p(f(c), e)}}
    {p(g(y_1), c)}
\]
\end{example}

\section{SAT-based Subsumption Resolution}%
\label{sec:sat-based-approach}

We describe the main steps of our SAT-based approach for deciding the applicability of subsumption resolution on a pair   $(L,M)$ of clauses.
The core of our work solves~\eqref{eq:subsumption-resolution-definition} by finding match substitutions between literals in $L$ and $M$. Our technique is summarised in Algorithm~\ref{alg:sat-subsumption-resolution}. 

\begin{algorithm}
  \caption{SAT-based subsumption resolution over  pair $(L,M)$ of clauses}
  \label{alg:sat-subsumption-resolution}
  \begin{algorithmic}
      \State $\ms \gets $ createMatchSet()
      \State $\solver \gets $ createSatSolver($\ms$)
      \Procedure{SubsumptionResolution}{$L,M$}
          \If {pruned($L,M$)}
              \State \Return{\NoSubsumptionResolution}
          \EndIf
          \If {fillMatchSet($\ms,L,M$) is  $\mathit{false}$}
              \State \Return{\NoSubsumptionResolution}
          \EndIf
          \State encodeConstraints($\solver,\ms$)
          \If {$\solver$.solve() is \SAT}
              \State \Return{buildConclusion($\solver$.getSolution(), $M$)} \Comment{conclusion of}
              \State \hfill {subsumption resolution}
          \EndIf
          \State \Return{\NoSubsumptionResolution}
      \EndProcedure
  \end{algorithmic}
\end{algorithm}

\paragraph{Pruning.}
The first step of Algorithm~\ref{alg:sat-subsumption-resolution} \emph{prunes} pairs  $(L,M)$ of clauses
that cannot be simplified by subsumption resolution due to a syntactic restriction over symbols in $L$ and $M$, \emph{viz.}\ whether the set of predicates in $L$ is a subset of the predicates in $M$. If not, then there is a literal in $L$ that cannot be matched to any literal in~$M$, and hence subsumption resolution cannot be applied.

\begin{example}\label{ex:SAT:pruning}
The clause pair $(L_4, M_4)$ from Figure~\ref{fig:Examples} is pruned by Algorithm~\ref{alg:sat-subsumption-resolution}: the set of predicates in $L_4$ and $M_4$ are respectively $\{p, q, r\}$ and $\{p, q\}$, implying that the literal $r(x_3)$ of $L_4$ cannot be matched to any literal in $M_4$.
\end{example}

\paragraph{Match set.} 
The \emph{match set} of Algorithm~\ref{alg:sat-subsumption-resolution} computes matching substitutions over literals of $L$ and $M$. The match set $\ms$ consists of a sparse matrix that assigns each literal pair $(l_i, m_j) \in L \times M$
a substitution $\sigma_{i,j}$ such that $\sigma_{i,j}(l_i) = m_j$ or $\sigma_{i,j}(l_i) = \neg m_j$. In addition, a polarity $P_{i,j}$ is also assigned to $(l_i, m_j)$, as follows: we set  polarity $P_{i,j} = +$ if  $\sigma_{i,j}(l_i) = m_j$ and $P_{i,j} = -$ if  $\sigma_{i,j}(l_i) = \neg m_j$.
This matrix is sparse because in general not all literal pairs $(l_i, m_j) \in L \times M$ can be matched.
Additionally, it is again possible to prune $(L,M)$
while filling the match set:
if a row of the match set is empty, then there is some literal in $L$ that cannot be matched to any literal in $M$.
In this case, subsumption resolution cannot use~$L$ to simplify~$M$, so the pair $(L,M)$ is pruned.

\paragraph{SAT solver.}
The $\solver$ of Algorithm~\ref{alg:sat-subsumption-resolution} is the CDCL-based SAT solver introduced previously~\cite{rath2022first},
which supports reasoning over matching substitutions in addition to standard propositional reasoning.
This solver also features direct support for \textit{AtMostOne} constraints. Solver performance was tuned for subsumption, which we retain for subsumption resolution.
Each propositional variable~$v$ is associated with a substitution~$\sigma_v$,
and the solver ensures that all substitutions~$\sigma_v$, for which~$v$ is assigned~$\top$ in the current model, are compatible.
Conceptually, a global substitution~$\sigma$ satisfying the invariant $\sigma = \bigcup \{\sigma_v\ |\ v = \top\}$
is kept in the SAT solver.
In the following, we will write this binding as $v\Rightarrow \sigma_v \subseteq \sigma$.

\begin{example}
    Suppose propositional variables $v_1$ and $v_2$ are associated with substitutions $\sigma_1 := \{ x \mapsto y \}$ and $\sigma_2 := \{ x \mapsto z \}$, respectively. 
    As $\sigma_1$ and $\sigma_2$ are incompatible, the solver will block assigning $v_1 = \top$ and $v_2 = \top$ simultaneously since it would break the above invariant.
\end{example}

\paragraph{Encoding constraints.}
Given the match set of $(L, M)$, we formalise the subsumption resolution problem~\eqref{eq:subsumption-resolution-definition} as the conjunction of four constraints over matching substitutions. Our  formalisation is given in Theorem~\ref{thm:subsumption-resolution-constraints} and is complete in the following sense: 
subsumption resolution can be applied over $(L,M)$ iff 
each constraint of Theorem~\ref{thm:subsumption-resolution-constraints} is satisfiable.
Application of subsumption resolution is tested via satisfiability checking over our constraints from Theorem~\ref{thm:subsumption-resolution-constraints}. Encodings of our subsumption resolution constraints are given in Section~\ref{sec:subsumption-resolution}. 

\paragraph{Building the conclusion.}
If a model is found for the constraints encoding subsumption resolution, 
the conclusion $M \setminus \{m'\}$ of \SR{} is built using the model.

\section{Subsumption Resolution and SAT Encodings}%
\label{sec:subsumption-resolution}

As mentioned in Section~\ref{sec:sat-based-approach}, we turn the application of subsumption resolution~\SR{} over~$(L, M)$ into the satisfiability checking problem of Algorithm~\ref{alg:sat-subsumption-resolution}.
We give our formalisation of~\SR{} in Theorem~\ref{thm:subsumption-resolution-constraints}, followed by two encodings to SAT (Section~\ref{sec:direct:SR}--\ref{sec:indirect:SR}) and adjustments to  subsumption  (Section~\ref{remark:subsumption-sat-constraints}). 

\begin{theorem}[Subsumption Resolution Constraints]
  Clauses~$M$ and~$L$ are the main and side premise, respectively, of an instance of the subsumption resolution rule \SR{}
  iff there exists a substitution $\sigma$ that satisfies the following four properties:
  \begin{align}
    &\textbf{existence} &
    \exists i\,j\ldotp\sigma(l_i) = \neg m_j \label{eq:subsumption-resolution-existence} \\
    &\textbf{uniqueness} &
    \exists j'\ldotp \forall i\,j\ldotp \bigl(\sigma(l_{i}) = \neg m_{j} \Rightarrow j = j' \bigr)
    \label{eq:subsumption-resolution-uniqueness} \\
    &\textbf{completeness} &
    \forall i \ldotp \exists j\ldotp \bigl( \sigma(l_i) = \neg m_j \lor \sigma(l_i) = m_j \bigr)
    \label{eq:subsumption-resolution-completeness}\\
    &\textbf{coherence} &
    \forall j \ldotp \bigl(\exists i \ldotp \sigma(l_i) = m_j\Rightarrow \forall i\ldotp \sigma(l_i) \neq \neg m_j\bigr)
    \label{eq:subsumption-resolution-coherence}
  \end{align}
  \label{thm:subsumption-resolution-constraints}
\end{theorem}

We relate these constraints to the definition of subsumption resolution~\eqref{eq:subsumption-resolution-definition}.
The \textbf{existence} property~\eqref{eq:subsumption-resolution-existence} requires a literal $m_j$ in $M$ such that a literal $l_i$ of $L$ can be matched to $\neg m_j$, 
ensuring the existence of the resolution literal in \SR.
\textbf{Uniqueness}~\eqref{eq:subsumption-resolution-uniqueness} asserts that the resolution literal $m_j$ of \SR{} is unique, required because \SR{} performs only a single resolution step.
\textbf{Completeness}~\eqref{eq:subsumption-resolution-completeness} requires each literal in $L$ be matched either to the complement of a resolution literal, or to a literal in $M$.  Since each (complementary) literal in $L$ is matched to one (resolution) literal of $M$, the completeness property ensures that the conclusion of \SR{}  subsumes $M$.
Finally, \textbf{coherence}~\eqref{eq:subsumption-resolution-coherence} states that all literals in $M$ must be matched by literals in $L$ with uniform polarity. This implies that all literals of $L$ other than the resolution literal are present in the conclusion of \SR.
We note that these constraints can be used to recreate Example~\ref{ex:SubsRes}.

\begin{example}\label{ex:uniqueness}
  The clause pair  $(L_2, M_2)$ of Figure~\ref{fig:Examples} does not satisfy the {uniqueness} property: both the match between $p(x_1)$ and $\neg p(y)$ and the match between $q(x_2)$ and $\neg q(c)$ are negative and so no substitution can satisfy all constraints simultaneously.
  Therefore, subsumption resolution cannot be applied over $(L_2,M_2)$.
\end{example}

\begin{example}\label{ex:coherence}
  The clause pair $(L_3,M_3)$ violates the coherence property for all possible $\sigma$, since a negative map from $p(x_1)$ to $\neg p(y)$ cannot coexist with a positive map from $\neg p(x_2)$ to $\neg p(y)$. Subsumption resolution cannot be performed over $(L_3,M_3)$.
\end{example}

\subsection{Direct SAT Encoding of Subsumption Resolution}\label{sec:direct:SR}
We present our encoding of subsumption resolution constraints as a SAT problem, allowing us to use Algorithm~\ref{alg:sat-subsumption-resolution} for deciding the application of \SR{}. 
In the sequel we consider the clauses $L,M$ as in Theorem \ref{thm:subsumption-resolution-constraints}. 

\paragraph{Compatibility.}
We introduce indexed propositional variables $b_{i, j}^+$ and $b_{i, j}^-$
to represent $\sigma(l_i) = m_j$ and $\sigma(l_i) = \neg m_j$ respectively,
which we use
to track compatible matching substitutions between literals of $L$ and $M$.
More precisely, a propositional variable is created if and only if the corresponding match is possible (i.e., in the formulas below, if no match exist, replace the corresponding propositional variable by $\bot$).
As it is not possible to have simultaneously a substitution $\sigma_{i, j}(l_i) = m_j$ and $\sigma_{i, j}(l_i) = \neg m_j$,
we also write $b_{i,j}$ to mean either $b_{i,j}^+$ or $b_{i,j}^-$ when the polarity of the match is irrelevant.
Following Section~\ref{sec:sat-based-approach}, the variables are bound to their substitutions:
\begin{align}
    &\textbf{SAT-based compatibility\qquad  }
  \bigwedge_{i}\bigwedge_{j} \left[b_{i,j} \Rightarrow \sigma_{i, j} \subseteq \sigma\right]
  \label{eq:subsumption-resolution-compatibility}
\end{align}

\paragraph{\SR{} constraints.} Constraints~\eqref{eq:subsumption-resolution-existence}--\eqref{eq:subsumption-resolution-coherence} of Theorem~\ref{thm:subsumption-resolution-constraints} employ \emph{bounded} quantification over the finite number of literals in $L,M$. 
Expanding these quantifiers over their respective domains, we translate them into the following SAT formulas: 
\begin{align}
  &\textbf{SAT-based existence  } &
  \bigvee_{i} \bigvee_{j} b_{i,j}^-
  \label{eq:subsumption-resolution-direct-existence} \\
  &\textbf{SAT-based uniqueness  } &
  \bigwedge_{j} \bigwedge_{i} \bigwedge_{i' \geq i}\bigwedge_{j'> j}\neg b_{i,j}^- \lor \neg b_{i',j'}^-
  \label{eq:subsumption-resolution-direct-uniqueness} \\
  &\textbf{SAT-based completeness  } &
  \bigwedge_{i} \bigvee_{j} b_{i,j}
  \label{eq:subsumption-resolution-direct-completeness}\\
  &\textbf{SAT-based coherence  } &
  \bigwedge_{j}\bigwedge_{i}\bigwedge_{i'}\neg b_{i,j}^+ \lor \neg b_{i',j}^-
  \label{eq:subsumption-resolution-direct-coherence}
\end{align}

\paragraph{\SR{} as SAT problem.}
Based on the above, application of subsumption resolution is decided by the satisfiability of
$\eqref{eq:subsumption-resolution-compatibility}\wedge\eqref{eq:subsumption-resolution-direct-existence}\wedge\eqref{eq:subsumption-resolution-direct-uniqueness}\wedge\eqref{eq:subsumption-resolution-direct-completeness}\wedge\eqref{eq:subsumption-resolution-direct-coherence}$.
This SAT formula extended with substitutions represents the result of $encodeConstraint()$ in  Algorithm~\ref{alg:sat-subsumption-resolution} and is used further in Algorithm~\ref{alg:sat-subsumption-resolution-new}.
When this formula is satisfiable, we construct the substitution $\sigma$ required for \SR{} by
$$\sigma = \bigcup \{\sigma_{i, j}\ |\ b_{i,j} = \top\}.$$
From the model of the SAT solver, we extract the 
first literal $b_{i,j}^-$ assigned $\top$, from which we conclude that the $j$\textsuperscript{th}
 literal in $M$ is the resolution literal of \SR{}. As such, application of \SR{} over $L$ and $M$ results in replacing $M$ by $M \setminus \{m_j\}$.

\begin{remark}
  Implicitly, all $l_i$ literals are mapped to at most one literal $m_j$. Indeed, if there were several literals $m_j$ such that $\sigma(l_i) = m_j$ or $\sigma(l_i) = \neg m_j$, then either the respective matches are not compatible (guarded by the {compatibility} property~\eqref{eq:subsumption-resolution-compatibility}), there are identical literals in $M$, or $M$ is a tautology (which is not allowed).
  \label{remark:subsumption-resolution-direct-uniqueness}
\end{remark}

\begin{remark}\label{remark:subsumption-resolution-direct-coherence}
While we defined $b_{i,j}$ to be true if, and \emph{only} if, $\sigma_{i, j} \subseteq \sigma$, we only encode the sufficient condition $b_{i,j} \Rightarrow \sigma_{i,j}\subseteq\sigma$. The {completeness} property~\eqref{eq:subsumption-resolution-direct-completeness} together with Remark \ref{remark:subsumption-resolution-direct-uniqueness} state that each $l_i$ must have exactly one match to some $m_j$ or $\neg m_j$. Therefore, if $\sigma_{i,j}\subseteq \sigma$ then the respective $b_{i,j}$ must be true and the condition also becomes necessary: $b_{i,j} \Leftarrow \sigma_{i,j}\subseteq \sigma$.
\end{remark}

\begin{example}\label{ex:SAT:direct}
Consider the pair $(L_1,M_1)$ of Figure~\ref{fig:Examples}. The match set $ms$ of Algorithm~\ref{alg:sat-subsumption-resolution} is:
\begin{equation*}
  \sigma_{i, j} =
  \left[
  \begin{matrix*}
    \{x_1 \mapsto g(y_1), x_2\mapsto c\} & \{x_1 \mapsto f(c), x_2\mapsto e\} \\
    \bot & \{x_1 \mapsto c, x_2\mapsto e\}
  \end{matrix*}
  \right]
  \ \ \
  P_{i, j} =
  \left[
  \begin{matrix*}
    + & - \\
      & -
  \end{matrix*}
  \right]
\end{equation*}
Since $\sigma_{2,1}$ is incompatible with any substitution, $b_{2,1} = \bot$ need not be defined. This also allows to disregard SAT clauses that are trivially satisfied. The {existence}~\eqref{eq:subsumption-resolution-direct-existence} and {completeness}~\eqref{eq:subsumption-resolution-direct-completeness} properties cannot have empty clauses: this is easily detected while filling the match set, and the instance of \SR\ is pruned. Adding falsified literals in these constraints is unnecessary. The {uniqueness}~\eqref{eq:subsumption-resolution-direct-uniqueness} and {coherence}~\eqref{eq:subsumption-resolution-direct-coherence} properties have only negative polarity literals and therefore there is no need to add clauses containing $b_{2,1}$.
In light of the previous comment, we use variables $b_{1,1}^+$, $b_{1,2}^-$ and $b_{2,2}^-$ and encode \SR{} using the following constraints: 
\begin{align*}
  & b_{1,1}^+  \Rightarrow  \{x_1 \mapsto g(y_1), x_2 \mapsto c\}\subseteq \sigma & \textbf{SAT-based compatibility} \textnormal{ of } b_{1,1}^+\\
  & b_{1,2}^-  \Rightarrow  \{x_1 \mapsto f(c), x_2 \mapsto e\}\subseteq \sigma & \textbf{SAT-based compatibility} \textnormal{ of } b_{1,2}^-\\
  & b_{2,2}^-  \Rightarrow  \{x_2 \mapsto c, x_3 \mapsto e\}\subseteq \sigma & \textbf{SAT-based compatibility} \textnormal{ of } b_{2,2}^-\\
  & b_{1,2}^- \lor b_{2,2}^- & \textbf{SAT-based existence} \\
  & b_{1,1}^+ \lor b_{1,2}^-& \textbf{SAT-based completeness}, i=1 \\
  & b_{2,2}^-& \textbf{SAT-based completeness}, i=2 \\
\end{align*}
The {uniqueness}~\eqref{eq:subsumption-resolution-direct-uniqueness}
and {coherence}~\eqref{eq:subsumption-resolution-direct-coherence} properties are
trivial here because the problem is simple: all $b_{i,j}^-$ have the same $j$, and no literal $m_j$ can be mapped with different polarities.
By using SAT solving from Algorithm~\ref{alg:sat-subsumption-resolution} over the above SAT constraints, we obtain the SAT model $b_{1,1}^+ \land \neg b_{1,2}^- \land b_{2,2}^-$, with $b_{2,2}^-$ the first literal assigned $\top$ with negative polarity. The application of \SR{} over $(L_1, M_1)$ yields the conclusion $M \setminus \{m_2\} = p(g(y_1), c)$, replacing $M$. 
\end{example}

\subsection{Indirect SAT Encoding of Subsumption Resolution}\label{sec:indirect:SR}
SAT-based formulas \eqref{eq:subsumption-resolution-direct-uniqueness} and \eqref{eq:subsumption-resolution-direct-coherence} may yield many constraints, with worst-case complexity $O(|L|^2 |M|^2)$. In practice such situations  rarely occur, since the match set $ms$ is  sparsely populated.
Nevertheless, to alleviate this worst-case complexity, we further constrain the approach of Section~\ref{sec:direct:SR}.
We introduce structuring propositional variables $c_j$ such that   $c_j$ is $\top$ iff there exists a literal $l_i$ with  $\sigma(l_i) = \neg m_j$,
which we encode as:

\begin{align}
  &\textbf{SAT-based structurality } &
  \bigwedge_j \left[\neg c_j \lor \bigvee_i b_{i,j}^- \right] \land \bigwedge_j \bigwedge_i \left(c_j \lor \neg b_{i,j}^-\right)
  \label{eq:subsumption-resolution-indirect-encoding}
\end{align}

\paragraph{\SR{} as revised SAT problem.}  While  the compatibility property~\eqref{eq:subsumption-resolution-compatibility} remains unchanged, the \SR{} constrains of Theorem~\ref{thm:subsumption-resolution-constraints} are revised as given below.  
\begin{align}
  &\textbf{SAT-based revised existence  } &
  \bigvee_{j} c_j
  \label{eq:subsumption-resolution-indirect-existence} \\
  &\textbf{SAT-based revised uniqueness  } &
  \mathit{AtMostOne}(\{c_{j}, j =1,...,|M|\})
  \label{eq:subsumption-resolution-indirect-uniqueness} \\
  &\textbf{SAT-based revised completeness  } &
  \bigwedge_{i} \bigvee_{j} b_{i,j}
  \label{eq:subsumption-resolution-indirect-completeness}\\
  &\textbf{SAT-based revised coherence  } &
  \bigwedge_{j} \bigwedge_{i} \left(\neg c_j \lor \neg b_{i,j}^+\right)
  \label{eq:subsumption-resolution-indirect-coherence}
\end{align}
Similarly to Section~\ref{sec:direct:SR}, application of subsumption resolution is decided via  Algorithm~\ref{alg:sat-subsumption-resolution} by checking satisfiability of
$\eqref{eq:subsumption-resolution-compatibility}\wedge
\eqref{eq:subsumption-resolution-indirect-encoding} \wedge
\eqref{eq:subsumption-resolution-indirect-existence}
\wedge
\eqref{eq:subsumption-resolution-indirect-uniqueness}
\wedge
\eqref{eq:subsumption-resolution-indirect-completeness}
\wedge
\eqref{eq:subsumption-resolution-indirect-coherence}
$.
Using the above SAT formula as the  result of $encodeConstraint()$ in  Algorithm~\ref{alg:sat-subsumption-resolution}, the worst-case behaviour is eliminated in exchange for $O(|M|)$ propositional variables, $c_j$. 
While the direct encoding of Section~\ref{sec:direct:SR} is more efficient  on small problems as  it requires fewer variables and constraints, the indirect encoding of this section is expected to behave better on larger problems (see Section~\ref{sec:experiments}).

\begin{remark}
  Note that the {uniqueness} property~\eqref{eq:subsumption-resolution-indirect-uniqueness} is handled via \textit{AtMostOne} constraints, based on the approach of~\cite{rath2022first}. If a variable $c_j$ is set to $\top$, then our SAT  $solver$  in Algorithm~\ref{alg:sat-subsumption-resolution} infers that all  other variables $c_{j'}$ are set to $\bot$.
\end{remark}

\begin{example}\label{ex:SAT:indirect}
Consider again the clause pair $(L_1, M_1)$ of Figure~\ref{fig:Examples}. Compared to Example~\ref{ex:SAT:direct}, our revised encoding of \SR{} requires one additional variable $c_2$,  as  $m_2$ in Example~\ref{ex:SAT:direct} is used with negative polarity. The revised constraints are:
\begin{align*}
  & b_{1,1}^+  \Rightarrow  \{x_1 \mapsto g(y_1), x_2 \mapsto c\} \subseteq \sigma & \textbf{SAT-based compatibility} \textnormal{ of } b_{1,1}^+\\
  & b_{1,2}^-  \Rightarrow  \{x_1 \mapsto f(c), x_2 \mapsto e\} \subseteq \sigma & \textbf{SAT-based compatibility} \textnormal{ of } b_{1,2}^-\\
  & b_{2,2}^-  \Rightarrow  \{x_2 \mapsto c, x_3 \mapsto e\} \subseteq \sigma & \textbf{SAT-based compatibility} \textnormal{ of } b_{2,2}^-\\
  & \neg c_{2} \lor b_{1,2}^- \lor b_{2,2}^- & \textbf{SAT-based structurality} \textnormal{ of } c_2\\
  & c_2 \lor \neg b_{1,2}^-& \textbf{SAT-based structurality}\textnormal{ of } c_2\\
  & c_2 \lor \neg b_{2,2}^-& \textbf{SAT-based structurality}\textnormal{ of } c_2\\
  & c_{2} & \textbf{SAT-based revised existence} \\
  & \mathit{AtMostOne}(\{ c_{2}\}) & \textbf{SAT-based revised uniqueness} \\
  & b_{1,1}^+ \lor b_{1,2}^-& \textbf{SAT-based revised completeness}, i=1 \\
  & b_{2,2}^-& \textbf{SAT-based revised completeness}, i=2 \\
\end{align*}
The SAT solver returns  $b_{1, 1}^+\land \neg b_{1, 2}^-\land b_{2, 2}^- \land c_2$ as a solution to the above SAT problem, from which the application of \SR{} yields a similar result to that of Example~\ref{ex:SAT:direct}. 
\end{example}

\newcommand{\eql}{\simeq}
\begin{remark}
    We note that our method naturally supports commutative predicates, such as equality. 
    Let $\eql$ denote object-level equality.
    Suppose we have literals $l_i := a \eql b$ and $m_j := c \eql d$.
    Two propositional variables with associated matching substitutions $\sigma_{i,j}$ and $\sigma'_{i,j}$ are introduced,
    where $\sigma_{i,j}$ matches $a \eql b$ against $c \eql d$
    and $\sigma'_{i,j}$ matches $a \eql b$ against $d \eql c$.
    If zero or one matches exist, then the problem behaves exactly like the non-symmetric case.
    If both matches exist, then $\sigma_{i,j}$ and $\sigma_{i,j}'$ must be incompatible:
    otherwise, $c$ and $d$ would be identical terms and the trivial literal~$m_j$ would have been eliminated.
    Therefore, our SAT-based encodings for subsumption resolution
    do not need to be adapted and behave as expected.
\end{remark}

\subsection{SAT Constraints for Subsumption}
\label{remark:subsumption-sat-constraints}
In the new framework of Algorithm \ref{alg:sat-subsumption-resolution}, the formulation suggested by~\cite{rath2022first} was adjusted to work with subsumption resolution.
Algorithm \ref{alg:sat-subsumption-resolution} needs very little adaptation for subsumption: the $\mathit{encodeConstraint}()$ method uses the encoding below,
and the conclusion needs not be built as only the satisfiability of the formulas is relevant.
The re-written SAT encoding becomes:
\begin{align}
  &\textbf{subsumption compatibility  } &
  \bigwedge_{i}\bigwedge_{j} \left(b_{i,j}^+ \Rightarrow \sigma_{i,j} \subseteq \sigma \right)
  \label{eq:subsumption-direct-encoding} \\
  &\textbf{subsumption completeness  } &
  \bigwedge_{i} \bigvee_{j} b_{i,j}^+
  \label{eq:subsumption-completeness}\\
  &\textbf{multiplicity conservation  } &
  \bigwedge_j \mathit{AtMostOne}(\{b_{i,j}^+, i = 1,...,|L|\})
  \label{eq:subsumption-multiplicity}
\end{align}

Note that the  set of propositional variables used in our SAT-based formulas \eqref{eq:subsumption-direct-encoding}--\eqref{eq:subsumption-multiplicity} encoding subsumption is a subset of the variables used by our SAT-based subsumption resolution constraints.

\paragraph{Pruning for subsumption.}
The pruning technique described in Section \ref{sec:sat-based-approach} can be adapted into a stronger form for subsumption. In this case, we will check for multi-set inclusion between multi-sets of (predicates, polarity) pairs.

\section{SAT-based Subsumption Resolution in Saturation}%
\label{sec:subsumption-resolution-in-vampire}

In this section we discuss the integration of our SAT-based subsumption resolution approach within saturation-based proof search. 

\paragraph{Forward/backward simplifications.}
For the purpose of efficient reasoning, saturation algorithms use two main variants of simplification inferences implementing redundancy.
\emph{Forward} simplifications are applied on a newly generated clause $M$ to check whether $M$ can be simplified by an existing clause $L$.
\emph{Backward} simplifications use a  newly generated clause $L$ to check whether $L$ can simplify existing clauses $M$. 
Backward simplification tends to be more expensive.

\paragraph{SAT-based subsumption resolution in saturation.}
Since subsumption is a stronger form of simplification, subsumption  is  checked before subsumption resolution. This means that  subsumption resolution is applied only if subsumption fails for all candidate premises.
We integrate Algorithm~\ref{alg:sat-subsumption-resolution} within saturation so that it is used both for subsumption and subsumption resolution.

\begin{algorithm}[t]
  \caption{SAT-based subsumption in saturation}
  \label{alg:sat-subsumption}
  \begin{algorithmic}
      \State $\ms \gets $ createMatchSet()
      \State $\solver \gets $ createSatSolver($\ms$)
      \Procedure{Subsumption}{$L,M$}
          \State $F_{\texttt{S}}$, $F_{\SR{}}$ $\gets$ pruned($L,M$) 
          \State \Comment{$F_{\texttt{S}}$ (resp. $F_{\SR{}}$) gets true if subsumption (resp. subsumption resolution) cannot succeed}
          \State fillMatchSet($\ms,L,M$) \Comment{Build the whole match set, and update $F_{\texttt{S}}$ and $F_{\SR{}}$}
          \If {$F_{\texttt{S}}$} \Comment{subsumption cannot be applied}
              \State \Return{\NoSubsumption}
          \EndIf
          \State encodeConstraints($\solver, \ms$) \Comment{SAT-constraints of Section~\ref{remark:subsumption-sat-constraints}}
          \If {$\solver$.solve() is \SAT}
            \State \Return{\Subsumption}
         \Else
             \State \Return{\NoSubsumption}
          \EndIf
      \EndProcedure
  \end{algorithmic}
\end{algorithm}
\begin{algorithm}[t]
  \caption{SAT-based subsumption resolution in saturation  \\
  {\color{white} for bit empty space only to get}-- with subsumption already set up via Algorithm~\ref{alg:sat-subsumption}}
  \label{alg:sat-subsumption-resolution-new}
  \begin{algorithmic}
    \Procedure{SubsumptionResolution}{$L,M$}
      \State \Comment{upon Algorithm~\ref{alg:sat-subsumption} failing to subsume}
      \State \Comment{the match set is already set up}
      \If {$F_{\SR{}}$}
        \State \Return{\NoSubsumptionResolution}
      \EndIf
      \State encodeConstraints($\solver,\ms$) \Comment{SAT constraints of Section~\ref{sec:direct:SR} or Section~\ref{sec:indirect:SR} }
      \If {$\solver$.solve() is \SAT}
        \State \Return{buildConclusion($\solver$.getSolution(), $M$)}\Comment{conclusion of}
        \State\hfill subsumption resolution
      \EndIf
      \State \Return{\NoSubsumptionResolution}
    \EndProcedure
  \end{algorithmic}
\end{algorithm}

Algorithms \ref{alg:sat-subsumption}--\ref{alg:sat-subsumption-resolution-new} display a variation of the integration of our SAT-based approach for checking subsumption resolution during saturation. Since most of the setup of subsumption is also required for subsumption resolution, both simplification rules are set up at the same time.
As such, whenever turning to subsumption resolution, the
same match set $ms$ from Algorithm~\ref{alg:sat-subsumption} can be reused, while also taking advantage of pruning steps performed during subsumption.

We modified the forward simplification algorithm as described in Algorithm~\ref{alg:forward-new}. In this new setting, checking the same pair $(L, M)$ for subsumption directly followed by subsumption resolution enables us to use Algorithms~\ref{alg:sat-subsumption}--\ref{alg:sat-subsumption-resolution-new} efficiently. 
Algorithm~\ref{alg:forward-new} pays the price of checking subsumption resolution even if subsumption may succeed, but in practice inefficiencies in this respect are seen rarely.

\begin{algorithm}[t]
  \centering
  \caption{Forward simplification with SAT-based subsumption resolution}
  \label{alg:forward-new}
  \begin{algorithmic}
      \Procedure{ForwardSimplify}{$M,F$}
        \State $M^* \gets$ \NoSubsumptionResolution
        \For{$L \in F \setminus \{M\}$}
            \If{subsumption($L, M$) is \Subsumption} \Comment{using Algorithm~\ref{alg:sat-subsumption}}
                \State $F \gets F \setminus \{M\}$
                \State \Return $\top$ \Comment{$M$ is subsumed and removed}
            \EndIf
            \If{$M^*$ = \NoSubsumptionResolution}
              \State $M^* \gets$ subsumptionResolution($L, M$) \Comment{using Algorithm~\ref{alg:sat-subsumption-resolution-new}}
            \EndIf
        \EndFor
        \If{$M^* \neq \NoSubsumptionResolution$}
          \State $F \gets F \setminus \{M\}\ \cup\ \{M^*\}$ \Comment{$M^*$ is the conclusion of subsumption resolution between $L$ and $M$}
          \State \Return $\top$
        \EndIf
        \State \Return $\bot$
      \EndProcedure
  \end{algorithmic}
\end{algorithm}

\paragraph{Role of indices.}
When applying inferences that require terms or literals to unify or match, modern automated first-order theorem provers typically use \emph{term indices}~\cite{term-indexing} to consider only viable candidates within the set of clauses.
Subsumption and subsumption resolution is no exception.
Our testbed system \vampire{} currently uses a substitution tree to index clauses for matching by their literals (Section~\ref{sec:experiments}).

\section{Implementation and Experiments}%
\label{sec:experiments}

\begin{algorithm}[t]
  \caption{Evaluation of SAT-based subsumption resolution}
  \label{alg:evaluation}
  \begin{algorithmic}
    \Procedure{ForwardSimplifyWrapper}{$M, F$}
      \State $s \gets \mathrm{startTimer}()$
      \State $r \gets \mathrm{ForwardSimplify}(M, F)$ \Comment{Benchmarked method}
      \State \algorithmiccomment{Prevent modification of $F$}
      \State $e \gets \mathrm{endTimer}()$
      \State $\mathrm{writeInFile}(e - s)$
      \State $r' \gets \mathrm{Oracle}(M, F)$
      \State $\mathrm{checkCoherence}(r, r')$  \algorithmiccomment{Empiric check}
      \State \Return{$r'$}
    \EndProcedure
  \end{algorithmic}
\end{algorithm}
We implemented and integrated our SAT-based subsumption resolution approach in the saturation-based first-order theorem prover \textsc{Vampire}~\cite{DBLP:conf/cav/KovacsV13}\footnote{\href{https://github.com/vprover/vampire/tree/robin_c-subsumption_resolution}{https://github.com/vprover/vampire/tree/robin\_c-subsumption\_resolution}}.

\paragraph{Versions compared.} 
We use following versions of \textsc{Vampire}  in our evaluation: 
\begin{compactitem}[$\bullet$\leftmargin=0em]
    \item \textsc{Vampire}$_M$ is the \emph{master} branch  without SAT-based subsumption resolution;
    \item \textsc{Vampire}$_I$ is the SAT-based subsumption resolution with the \emph{indirect} encoding  of Section~\ref{sec:indirect:SR} and a standard forward simplification
    algorithm with Algorithm \ref{alg:sat-subsumption-resolution} --- that is, Algorithm \ref{alg:forward-new} is not used here;
    \item \textsc{Vampire}$^*_I$ uses the \emph{indirect} encoding with Algorithms \ref{alg:sat-subsumption}--\ref{alg:forward-new};
    \item \textsc{Vampire}$^*_D$ uses the \emph{direct} encoding of Section~\ref{sec:direct:SR} and Algorithms \ref{alg:sat-subsumption}--\ref{alg:forward-new}.
\end{compactitem}

\paragraph{Experimental setting.}
To evaluate our work, we used the examples of the TPTP library (version 8.1.2)~\cite{Sut17}.
In our evaluation, 24\,926 problems were used out of the 25\,257 TPTP problems;
the remaining problems are not supported by \textsc{Vampire} (e.g., problems with both higher-order operators and polymorphism).

Our experimental evaluation was done on a machine with two 32-core AMD Epyc 7502 CPUs clocked at 2.5\,GHz
and 1006\,GiB of RAM (split into 8 memory nodes of 126\,GiB shared by 8 cores). 
Each benchmark problem was run with the options \texttt{-sa otter -t 60}, meaning that we used the \textsc{Otter} saturation algorithm~\cite{otter} with a  60-second time-out. We use the \textsc{Otter} strategy because it is the most aggressive in terms of simplification and therefore runs the most subsumption resolutions.  We  turned off the AVATAR framework  (\texttt{-av off}) in order to have full control over SAT-based reasoning in \textsc{Vampire}.

\paragraph{Evaluation setup.}
Our evaluation process is summarised in Algorithm~\ref{alg:evaluation}, incorporating the following notes. 

\begin{compactitem}[$\bullet$\leftmargin=0em]
  \item The conclusion clause of the subsumption resolution rule \SR{} is not necessarily unique. Therefore,  different versions of subsumption resolution, including
  our work based on direct and indirect SAT encodings, may not  return the same conclusion clause of \SR{}. Hence, applying different versions of subsumption resolution over the same clauses  may change the saturation  process.
  \item Saturation with our SAT-based subsumption resolution  takes advantage of  subsumption checking (see Algorithms~\ref{alg:sat-subsumption-resolution-new}--\ref{alg:forward-new}). Therefore, only  checking subsumption resolution  on pairs of clauses is not a fair nor viable comparison, as isolating subsumption checks from subsumption resolution is not what we aimed for (due to efficiency). 
  \item CPU cache influences results. For example, two consecutive runs of  Algorithm~\ref{alg:forward-new} may be up to 25\% faster on second execution, due to cache effects. 
\end{compactitem}

For the reasons above, we decided to measure the run time of a complete execution of Algorithm~\ref{alg:forward-new}. To prevent the branches to change, an \texttt{Oracle} is used to choose the path to follow. The \texttt{Oracle} is  based on our indirect SAT encoding (\textsc{Vampire}$^*_{I}$). This way, the same computation graph is used for all  evaluated methods.
To prevent cache preheating, we run the \texttt{Oracle} after the respective evaluated method. This way the cache is in a normal state for the evaluated method. 
To measure the  run time of Algorithm~\ref{alg:forward-new}, a \texttt{Wrapper} method was built on top of the \texttt{Forward Simplify} procedure of Algorithm~\ref{alg:forward-new}. This  \texttt{Wrapper} replaces the \texttt{Forward Simplify} loop in \textsc{Vampire} with minimal changes to the code. 
To empirically verify the correctness of our results, we used the \texttt{Wrapper} to compare the result of the evaluated method with the result of the \texttt{Oracle}. 

\begin{figure}[t]
  \centering
  \includegraphics[scale=0.5]{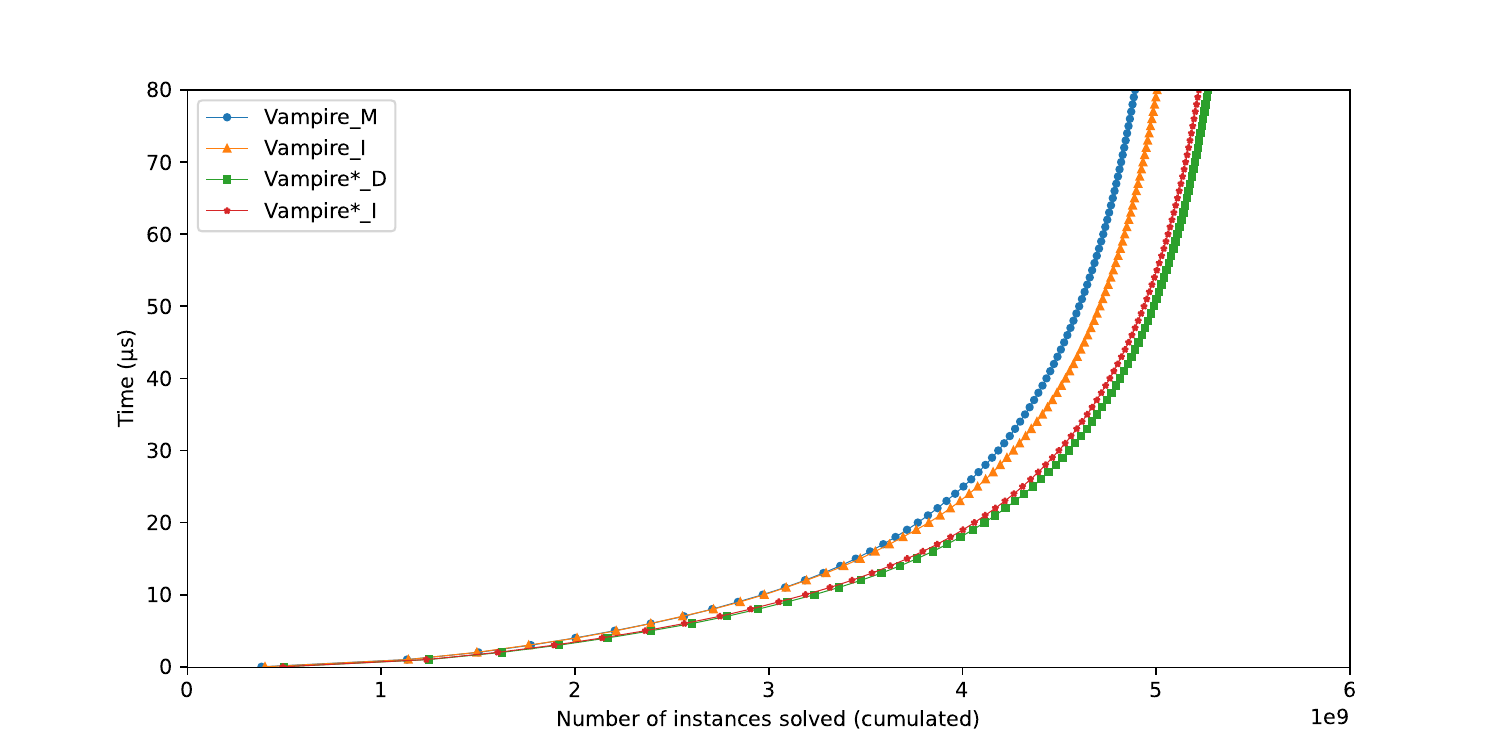}
  \caption{%
    Cumulative instances of applying subsumption resolution, using  the TPTP examples.
    A point $(n,t)$ on the graph means that $n$
    forward simplify loops were executed in less than $t~\mu s$.
    The flatter the curve, the faster the \textsc{Vampire} version is.
    }
  \label{fig:cummulative-graph}
\end{figure}

\paragraph{Experimental details and analysis.} 
Figure \ref{fig:cummulative-graph} lists the cumulative instances solved by the respective \textsc{Vampire} versions, highlighting the strength of forward simplifications for effective saturation.

\begin{remark}
Our experimental summary in Figure \ref{fig:cummulative-graph} shows that the total number of \texttt{Forward Simplify} loops ran in 60 seconds. However, the average and standard deviation were computed only on the intersection of the problems solved. That is, only the \texttt{Forward Simplify} loops finished by all the methods are taken into account. Otherwise, if a hard problem is solved in, for instance, 1\,000\,000 $\mu s$ by one method, and times out for another, the average for the better would increase a lot, but the weaker method would not be penalised. Table~\ref{tab:results} summarises the
average solving time of our evaluation.

\begin{table}[t]
  \setlength{\tabcolsep}{6pt}
  \centering
  \begin{tabular}{c | c r | r}
    \hline
    \textbf{Prover} & \textbf{Average} & \textbf{Std.\ Dev.} & \textbf{Speedup}\\
    \hline
    \textsc{Vampire}$_M$  & $42.63\,\mu s$ & $1609.06\,\mu s$ \quad & 0\,\% \quad \\
    \textsc{Vampire$_I$}                   & $40.13\,\mu s$ & $1554.52\,\mu s$ \quad & 6.2\,\% \quad \\
    \textsc{Vampire}$^*_{D}$ & $34.39\,\mu s$ & $1047.85\,\mu s$ \quad & 23.9\,\% \quad \\
    \textsc{Vampire}$^*_{I}$ & $34.55\,\mu s$ & $250.25\,\mu s$ \quad & 23.4\,\% \quad \\
    \hline
  \end{tabular}\vspace*{0.5em}
  \caption{Average time spent in the \texttt{Forward Simplify} loop. \textsc{Vampire}$^*_{D}$ is the fastest method, closely followed by the \textsc{Vampire}$^*_{I}$. However, the indirect encoding is much more stable and has a lower variance.}
  \label{tab:results}
\end{table}
\end{remark}

\begin{figure}[p]
  \centering
  \begin{subfigure}{1\textwidth}
    \centering
    \includegraphics[trim={5cm 0 9.5cm 0}, scale=0.3]{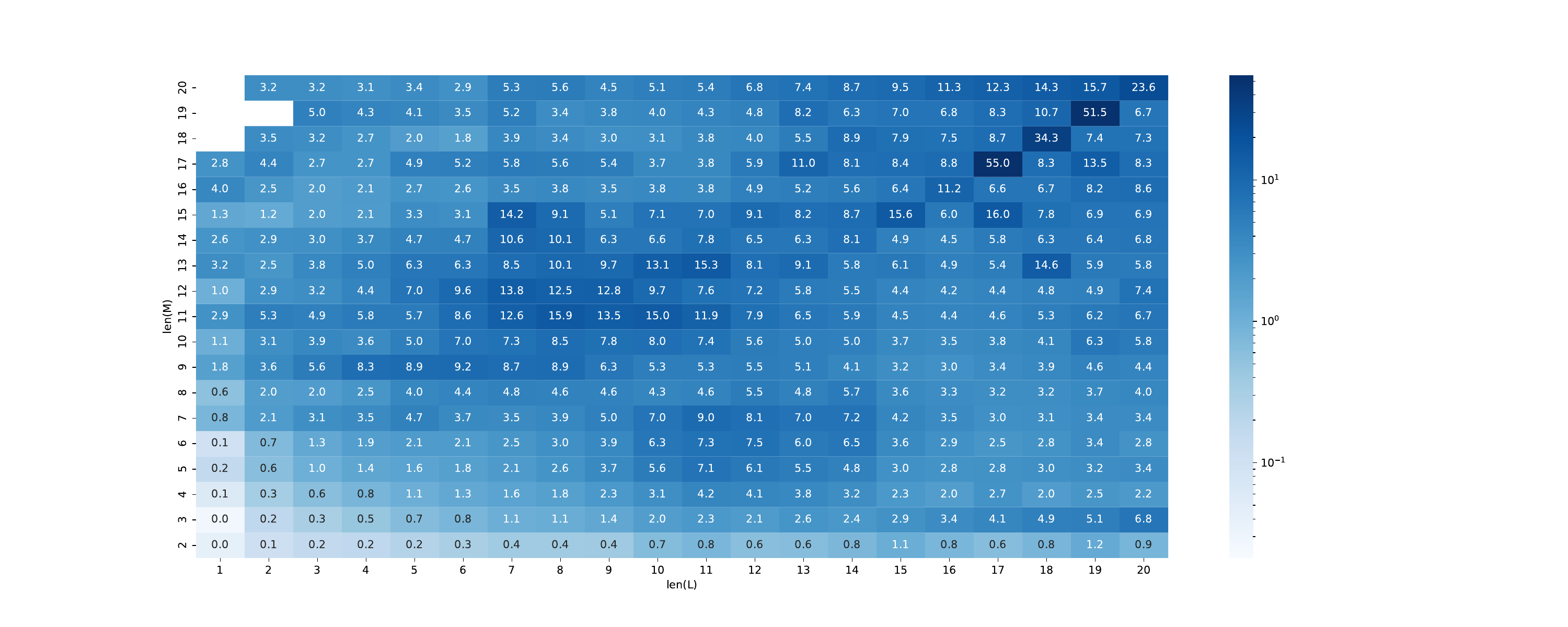}
    \caption{Average time ($\mu s$)  for creating/solving   {direct} encoding constraints (Section~\ref{sec:direct:SR}).}
    \label{fig:heatmap-averagetime_sat_1}
  \end{subfigure}
  \begin{subfigure}{1\textwidth}
    \centering
    \includegraphics[trim={5cm 0 9.5cm 0}, scale=0.3]{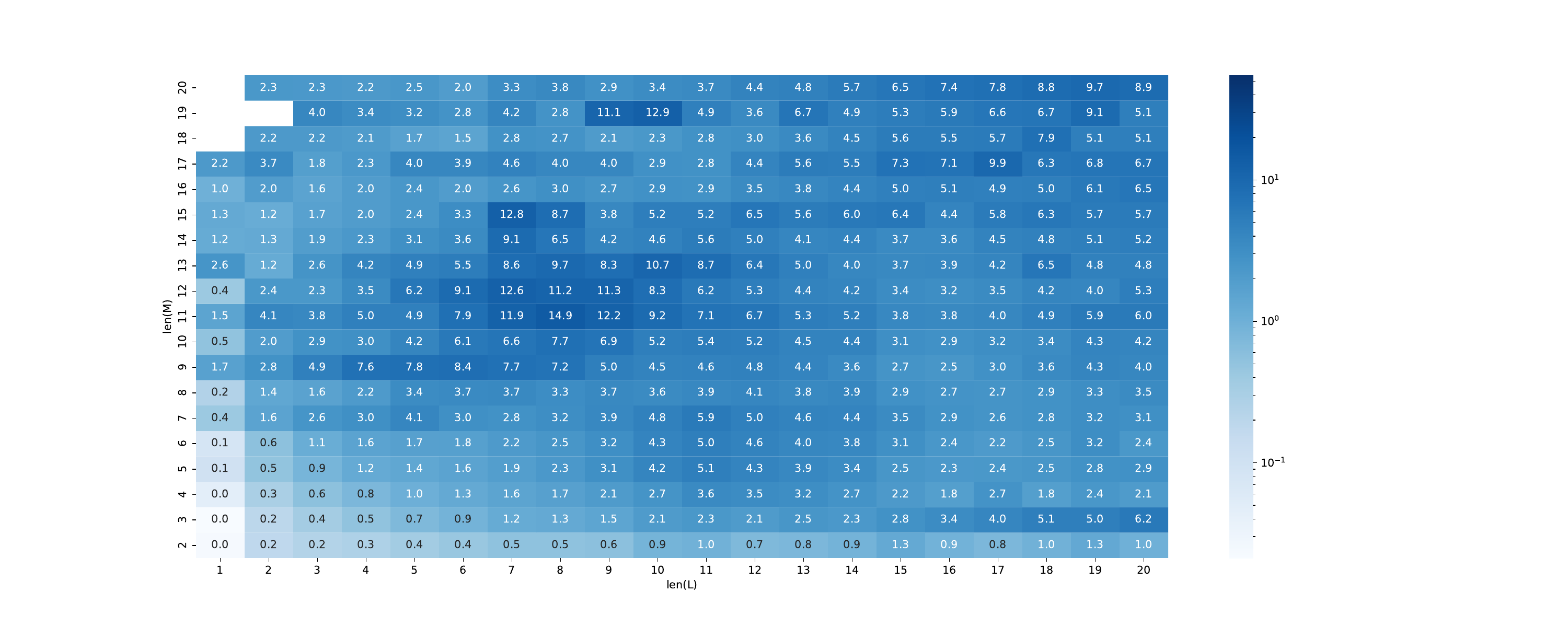}
    \caption{Average time ($\mu s$) for 
    creating/solving   {indirect} encoding constraints (Section~\ref{sec:indirect:SR}).}
    \label{fig:heatmap-averagetime_sat_2}
  \end{subfigure}
  \caption{Average time  ($\mu s$) spent on the creating and solving SAT-based subsumption resolution constraints.}
  \label{fig:heatmap-averagetime}
\end{figure}

\paragraph{Comparison of  encodings.}
We correlated the constraint building and SAT solving time with the length of clauses, using the different encodings of Sections~\ref{sec:direct:SR}--\ref{sec:indirect:SR}. Figure \ref{fig:heatmap-averagetime} shows that  on larger clauses, the average computation time increases faster for the direct encoding than for the indirect encoding.

\begin{table}[p]
    \setlength{\tabcolsep}{6pt}
    \centering
    \begin{tabular}{c|cc}
        \hline
        \textbf{Prover} & \textbf{Total Solved} & \textbf{Gain/Loss}
        \\
        \hline
        \textsc{Vampire}$_M$ & 10\,555 & baseline \\
        \vampire{}$^*_{D}$ & 10\,667 & ($+141$, $-29$) \\
        \vampire{}$^*_I$ & 10\,658 & ($+133$, $-30$) \\
        \hline
    \end{tabular}\vspace*{0.5em}
    \caption{%
        Number of TPTP problems solved by the considered versions of \textsc{Vampire}.
        The run was made using the options \texttt{-sa otter -av off} with a timeout of~60\,s.
        The \textbf{Gain/Loss} column reports the difference of solved instances compared to \vampire{}$_M$.
    }
    \label{tab:performances-vampire-vanilla}
\end{table}

\paragraph{Experimental summary.} 
Our experiments show that  
\vampire{}$^*_I$ yields the most stable approach for SAT-based subsumption resolution (Table~\ref{tab:results}), especially when it comes on solving large instances (Figure~\ref{fig:heatmap-averagetime}). Our results demonstrate the  superiority of SAT-based subsumption resolution used with forward simplifications in saturation (e.g., \vampire{}$^*_D$ and \vampire{}$^*_I$), as  concluded by  Table~\ref{tab:performances-vampire-vanilla}.

\section{Conclusion}%
\label{sec:conclusion}

We advocate SAT solving for improving saturation-based first-order theorem proving.
We encode powerful simplification rules, in particular subsumption resolution, as SAT problems, triggering eager and efficient reasoning steps for the purpose of keeping proof search small.
Our experiments with \vampire{} showcase the benefit of
SAT-based subsumption.
In the future, we aim to further extend simplification rules with SAT solving, in particular focusing on subsumption demodulation for equality reasoning~\cite{DBLP:conf/cade/GleissKR20}.

\paragraph{Acknowledgements.}
This work is the result of a research internship hosted at TU Wien and defended
at the University of Liège.
The authors would like to thank Pascal Fontaine for valuable discussions and
comments. We acknowledge funding from the ERC Consolidator Grant ARTIST
101002685 and the FWF SFB project SpyCoDe F8504.

\end{document}